\documentclass{article}

\usepackage[preprint]{neurips_2026}

\usepackage[utf8]{inputenc}
\usepackage[T1]{fontenc}
\usepackage{hyperref}
\usepackage{url}
\usepackage{booktabs}
\usepackage{amsfonts}
\usepackage{amsmath}
\usepackage{amssymb}
\usepackage{nicefrac}
\usepackage{microtype}
\usepackage{xcolor}
\usepackage{graphicx}
\usepackage{subcaption}
\usepackage{enumitem}
\usepackage{algorithm}
\usepackage{algpseudocode}
\usepackage{natbib}
\usepackage{multirow}
\usepackage{subcaption} 
\usepackage{wrapfig}
\usepackage{amsmath}
\usepackage{amsfonts}

\usepackage{amsthm}

\definecolor{canyon}{RGB}{205, 92, 92}

\title{GeoCycler: Reward-Aligned 3D Diffusion for Constraint-Conditioned Cyclic Peptide Design}

\author{
  Jingjie Zhang$^{1,*}$ \quad
  Hanqun Cao$^{1,*}$ \quad
  Haosen Shi$^{1}$ \quad
  Mutian He$^{4}$ \quad
  Yu Wang$^{5}$ \\
  Zijun Gao$^{1}$ \quad
  Fang Wu$^{6}$ \quad
  Xiaojun Yao$^{4}$ \quad
  Chang-Yu Hsieh$^{3}$ \quad
  Sinno Jialin Pan$^{1}$ \\
  Pranam Chatterjee$^{2}$ \quad
  Chunbin Gu$^{1}$ \quad
  Pheng-Ann Heng$^{1}$ \\
  \\
  $^{1}$The Chinese University of Hong Kong \quad
  $^{2}$University of Pennsylvania \quad
  $^{3}$Zhejiang University \\
  $^{4}$Macao Polytechnic University \quad
  $^{5}$Peking University \quad
  $^{6}$Stanford University \\
  \\
  {\small $^{*}$Equal contribution.}
}

\begin{document}

\maketitle
\begin{abstract}

Cyclic peptides are attractive therapeutic modalities because their closed-ring topology can improve stability and target specificity. However, de novo cyclic peptide design remains challenging for diffusion generators, as macrocyclization requires satisfying sparse, non-smooth, and compositional geometric constraints. Existing constraint-conditioned methods largely rely on inference-time guidance, which can steer samples toward desired closures but does not directly change the learned generative distribution. We propose \textbf{GeoCycler}, a reward-weighted diffusion alignment framework for training conditional latent diffusion models toward macrocyclization feasibility. GeoCycler introduces a type-gated stair reward that activates distance-based shaping only when prerequisite residue or linker types are satisfied, providing dense geometric feedback while avoiding misleading signals from chemically incompatible anchors. Together with positive-only reward weighting and replay-based stabilization, GeoCycler aligns a single generator across multiple cyclization topologies. On the LNR benchmark, GeoCycler improves pass@5 closure success over strong guidance-based baselines across stapled, head-to-tail, disulfide, and bicyclic settings. 
In particular, it improves head-to-tail success by 20.8 percentage points over CP-Composer while maintaining comparable amino-acid and backbone-dihedral statistics. These results suggest that training-time alignment to sparse geometric constraints is a promising alternative to relying solely on post hoc sampling-time correction for cyclic peptide generation.

\end{abstract}

\section{Introduction}
\label{sec:introduction}
Cyclic peptides provide an attractive molecular modality between small molecules and biologics. 
Their closed-ring architectures, formed through backbone cyclization or side-chain crosslinks, can restrict conformational flexibility, improve proteolytic stability, and pre-organize binding-competent structures \citep{zorzi2017cyclic, wang2022therapeutic, ji2024cyclic}. 
These properties make cyclic peptides promising candidates for targeting protein--protein interfaces and other challenging therapeutic sites \citep{ji2024cyclic, dougherty2019understanding}. 
However, the same closed-ring topology that makes cyclic peptides attractive also makes them difficult to generate: a valid candidate must satisfy both discrete chemical prerequisites, such as compatible anchor residues or linker motifs, and continuous geometric constraints, such as closure distances, local junction geometry, and coupled multi-loop arrangements.

Deep generative models have substantially expanded the design space for proteins and peptides, with diffusion, flow-based, and discrete models enabling scalable molecular and structural generation \citep{watson2023novo, kong2024full, tang2025peptune,tang2025tr2, chen2025areuredi,chen2025mogdfm}. 
Yet macrocyclization remains a distinct challenge. 
A model may generate plausible peptide-like backbones while still failing to place the correct anchors within a narrow closure geometry, especially when several closure edges must be satisfied simultaneously. 
This sparse and compositional structure makes cyclic peptide design different from optimizing many global structural properties: feasibility depends on whether local chemical types and local 3D geometry become compatible under a prescribed cyclization topology.

\begin{figure}[t]
    \centering
    \includegraphics[width=0.85\linewidth]{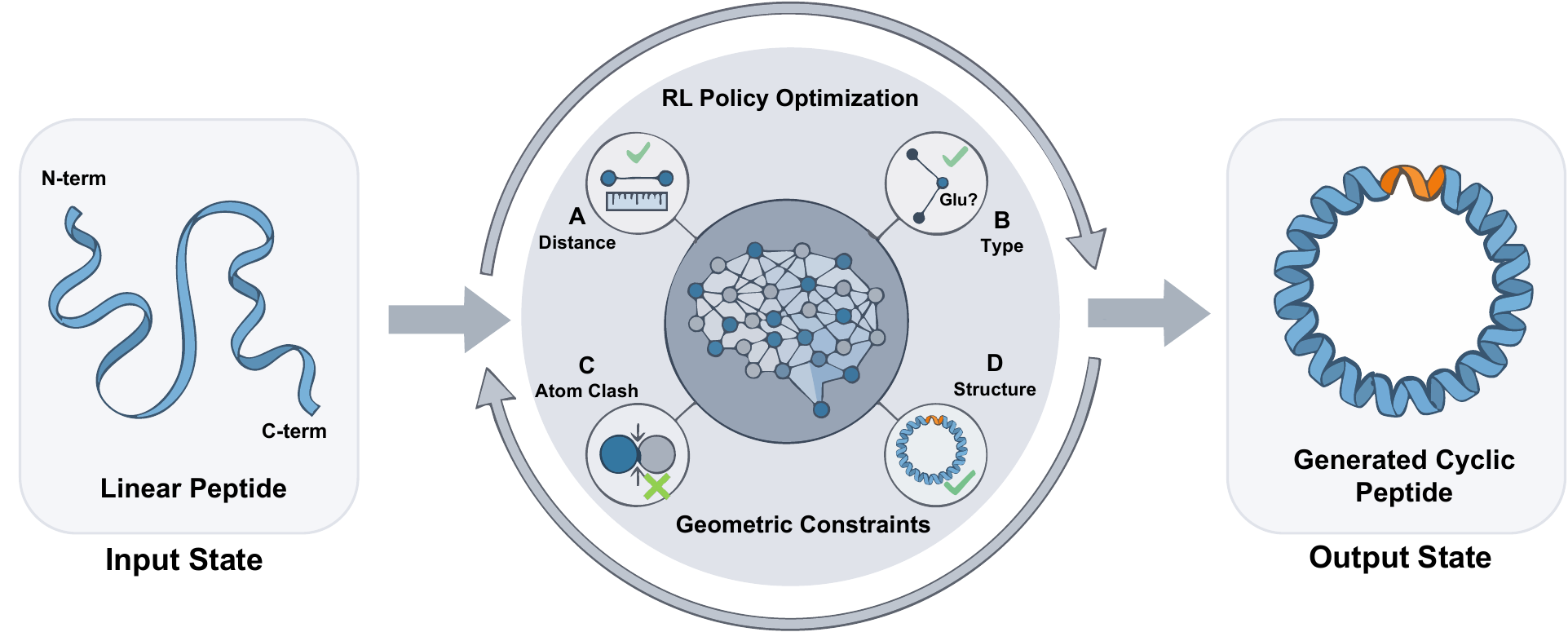} 
    \caption{\textbf{Cyclic peptide generation under hybrid geometric constraints.} 
    Cyclic peptide design requires satisfying both discrete prerequisite conditions, such as compatible anchor residues or linker motifs, and continuous closure-oriented geometry. 
    GeoCycler studies whether training-time policy alignment can increase the probability of generating closure-consistent candidates, while post-hoc structural screens are used to assess relaxation stability and severe steric artifacts.}
    \label{fig:intro}
\end{figure}

A widely used strategy for controllable generation is inference-time guidance. 
In cyclic peptide design, classifier-free guidance (CFG) \citep{ho2022classifier} can steer a pre-trained diffusion model toward samples that better satisfy user-specified cyclization constraints \citep{jiang2025zero}. 
However, this treats feasibility as an external correction during sampling rather than as a property learned by the generator. 
When the base distribution is only weakly biased toward cyclic closure, guidance must compensate for a mismatch between the learned prior and the desired constraint manifold, which becomes less reliable under sparse, coupled, or partially discrete constraints. 
This motivates a training-time alignment view: instead of only steering samples at inference time, we adapt the diffusion policy so that closure-consistent structures become more probable under its own generative distribution.

Cyclic peptide macrocyclization is a useful testbed for training-time alignment because its constraints are non-smooth, partially discrete, and compositional. 
Success is often defined by thresholded geometric criteria; distance feedback is meaningful only after prerequisite residue or linker types are present; and multi-loop topologies require several closures to be satisfied jointly. 
These properties make standard supervised fine-tuning poorly matched to the task and motivate policy-level optimization of diffusion generators under their own sampling distribution \citep{black2023training, fan2023dpok, wallace2024diffusion, zheng2025diffusionnft}.

We propose GeoCycler, a reward-aligned diffusion framework for cyclic peptide generation under compositional geometric constraints. 
The key idea is to move cyclization control from inference-time correction to training-time diffusion alignment. 
Building on recent advances in forward-process reinforcement learning for diffusion models \citep{zheng2025diffusionnft}, GeoCycler incorporates topology-dependent reward signals into the diffusion fine-tuning objective, allowing the generator to increase the likelihood of closure-consistent structures under its own sampling distribution. 
To make this alignment effective for 3D macrocyclization, GeoCycler introduces a type-gated geometric reward that couples chemical compatibility with bounded closure-distance shaping. 
The reward activates distance-based feedback only when the corresponding cyclization anchors are present, thereby avoiding misleading updates from chemically incompatible configurations while still providing informative signals for near-feasible closures. 
Rather than training separate models for different chemistries, GeoCycler learns a unified constraint-conditioned generator over four representative cyclic peptide topologies: stapled, head-to-tail, disulfide, and bicycle. 
On the LNR benchmark \citep{tsaban2022harnessing}, GeoCycler consistently improves closure success over guidance-based and reward-aligned baselines, with particularly strong gains on head-to-tail and disulfide cyclization. 
Additional no-steering and post-hoc structural analyses indicate that the improvement is not merely a consequence of stronger inference-time guidance, but reflects a shift in the aligned model distribution toward closure-consistent candidates.

Our contributions are summarized as follows:\begin{itemize}[leftmargin=*,nosep]\item \textbf{A training-time alignment framework for 3D diffusion under compositional constraints.}We recast cyclic peptide macrocyclization as a sparse, hybrid discrete-continuous alignment problem and adapt forward-process diffusion reinforcement learning to constraint-conditioned 3D structure generation.This shifts cyclization control from post-hoc sampling correction to optimization of the generator itself.

\item \textbf{A type-gated surrogate reward for stable macrocyclization learning.}
We design a topology-aware reward that separates prerequisite anchor compatibility from continuous closure geometry. 
By activating distance-based shaping only after the relevant type conditions are met, GeoCycler provides dense but chemically meaningful feedback for sparse ring-closure objectives.

\item \textbf{A unified generator for multiple cyclic peptide topologies.}
Rather than training topology-specific models, GeoCycler aligns a single conditional generator across stapled, head-to-tail, disulfide, and bicyclic constraints. 
On the LNR benchmark, this unified model improves closure-consistent generation over guidance-based and reward-aligned baselines, with additional analyses supporting distribution-level alignment beyond test-time steering.

\end{itemize}

\begin{figure*}[t]\centering\includegraphics[width=\linewidth]{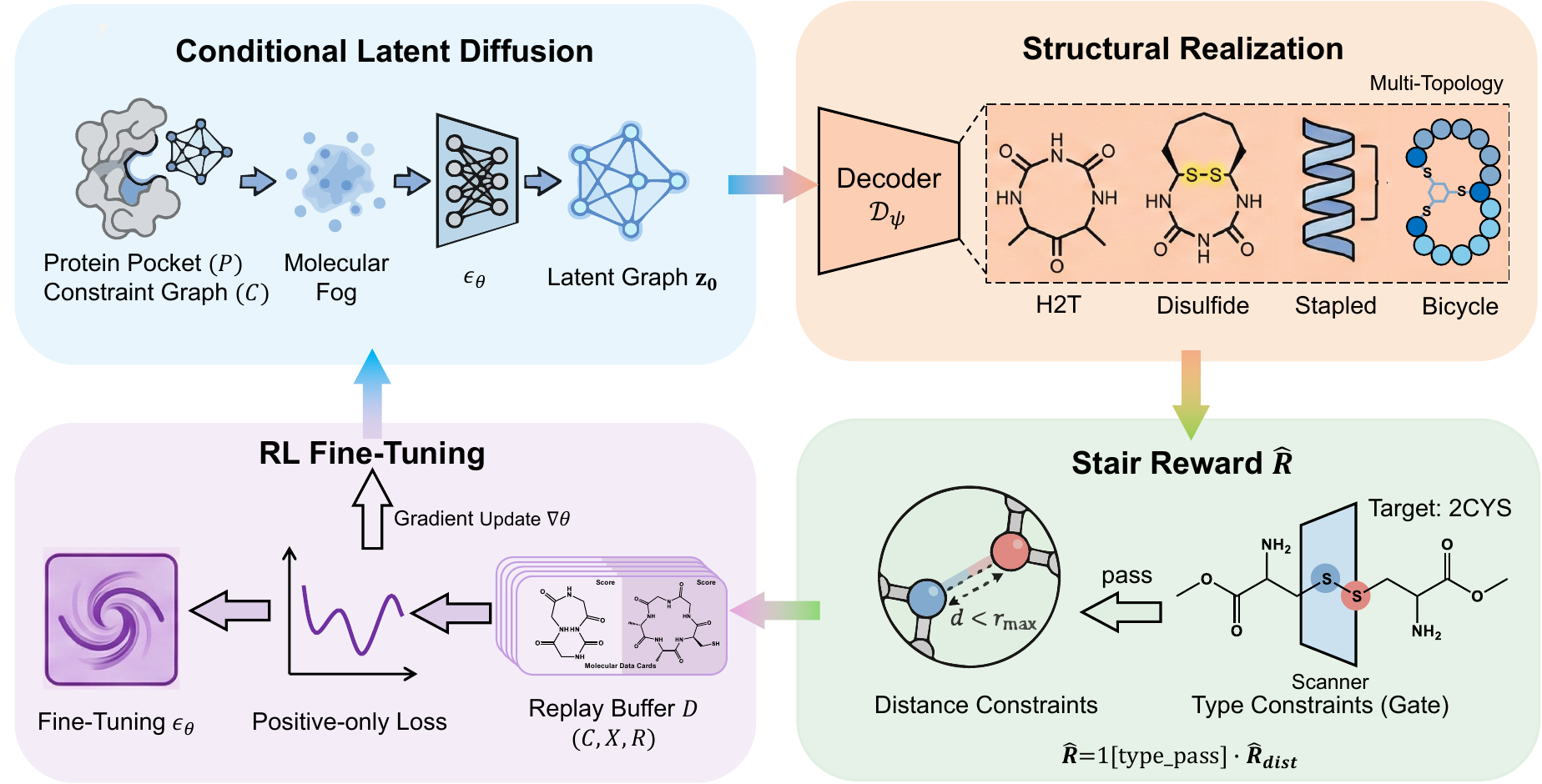}\caption{\textbf{Overview of GeoCycler.} GeoCycler fine-tunes a conditional latent diffusion generator through reward-weighted policy alignment. The framework combines type-gated geometric surrogate rewards, positive-only updates, and replay-based stabilization to increase the probability of closure-consistent cyclic peptide candidates across multiple topology constraints.}\label{fig}\end{figure*}

\section{Related works}
\label{sec:related}
\subsection{Generative Models for Cyclic Peptide Design}

Deep generative models have substantially advanced \textit{de novo} biomolecular design. 
3D equivariant diffusion models provide scalable samplers for geometric structure generation \citep{hoogeboom2022equivariant}, with RFdiffusion demonstrating strong conditional backbone design \citep{watson2023novo}. 
Peptide-specific models such as PepGLAD \citep{kong2024full} and PepFlow \citep{li2024full} further extend latent diffusion and flow matching to full-atom peptide design. 
However, generic peptide generators are not inherently optimized for macrocyclization, where candidates must satisfy both residue-type prerequisites and topology-specific closure geometry.

Recent cyclic peptide methods address this challenge more directly. 
Structure-prediction-based approaches adapt AlphaFold2-style models to circular topologies \citep{jumper2021highly, rettie2025cyclic}, while generative frameworks such as RFpeptides \citep{rettie2025accurate}, CPSDE \citep{zhou2025designing}, and CP-Composer \citep{jiang2025zero} incorporate macrocyclic constraints through specialized architectures, harmonic dynamics, atom-bond modeling, or composable geometric guidance. 
CP-Composer is particularly relevant to our setting because it supports multiple cyclization chemistries through constraint-conditioned generation and inference-time guidance. 
GeoCycler is complementary to these models: rather than introducing a new backbone generator, we study how a pretrained conditional diffusion model can be aligned at training time so that closure-consistent candidates become more likely under the learned distribution.

\subsection{Reward Alignment for Molecular and Diffusion Models}

Reinforcement learning has been widely used to optimize molecular and biological generators toward non-differentiable objectives, including binding, stability, activity, and fitness scores. 
Prior work includes search-based sequence design methods such as MCTS \citep{wang2023self,liu2025protinvtree,tang2025peptune}, policy-optimization approaches for biological sequences \citep{sun2025accelerating,xu2025protein,cao2025deep,cao2025glide,cao2025supervision}, and reward-driven structural design for protein assembly or complexes \citep{lutz2023top,gao2024deep}. 
These methods mainly optimize sequence-level or global functional objectives, whereas cyclic peptide macrocyclization requires satisfying sparse local geometric constraints coupled with discrete chemical prerequisites.

Recent work has extended reward alignment to diffusion and flow-based generators. 
DDPO \citep{black2023training} formulates diffusion sampling as a sequential decision process, while MOG-DFM \citep{chen2025mogdfm}, AReUReDi \citep{chen2025areuredi}, TR2 \citep{tang2025tr2}, and DiffusionNFT \citep{zheng2025diffusionnft} explore multi-objective, trajectory-aware, or forward-process fine-tuning strategies. 
GeoCycler brings this alignment perspective to continuous 3D cyclic peptide generation, where rewards are sparse, non-smooth, and topology-dependent. 
Compared with cyclic peptide RL/search methods such as HighPlay \citep{lin2025highplay} and CYC\_BUILDER \citep{wang2025reinforcement}, which search over sequence or assembly decisions for individual design problems, GeoCycler performs training-time reward alignment of a conditional 3D diffusion generator across multiple cyclization topologies.

\section{Methods}
\label{sec:method}
\subsection{Task Formulation}

We study target-conditioned cyclic peptide generation under prescribed macrocyclization constraints. 
Given a protein receptor or binding pocket $P$, the goal is to generate a peptide 
$\mathcal{M}=(\mathbf{X},\mathbf{S})$, where 
$\mathbf{X}\in\mathbb{R}^{L\times A\times 3}$ denotes peptide atom coordinates, 
$\mathbf{S}\in\mathcal{V}^{L}$ denotes the amino-acid sequence, $L$ is the peptide length, and 
$A$ denotes the number of atom slots represented per residue. 
Following CP-Composer~\citep{jiang2025zero}, we represent macrocyclization conditions as
$\mathcal{C}=(\mathcal{C}_{\mathrm{type}},\mathcal{C}_{\mathrm{dist}})$, where 
$\mathcal{C}_{\mathrm{type}}$ specifies prerequisite residue or linker patterns and 
$\mathcal{C}_{\mathrm{dist}}$ specifies topology-dependent geometric closure criteria. 
The conditional generator models
\begin{equation*}
p_{\theta}(\mathbf{X},\mathbf{S}\mid P,\mathcal{C}).
\end{equation*}

This formulation treats cyclic peptide design as a hybrid discrete-continuous constrained generation problem. 
The discrete component determines whether the generated sequence contains the required topology-specific anchor pattern, while the continuous component evaluates whether the corresponding anchor-proxy geometry satisfies the prescribed closure criterion. 
We consider four representative macrocyclization topologies: stapled, head-to-tail, disulfide, and bicyclic peptides. 
Detailed reward definitions and topology-specific criteria are provided in Appendix~\ref{app:reward}.

\subsection{Base Diffusion Policy}

GeoCycler is formulated as a training-time alignment framework for conditional 3D diffusion generators. 
In our experiments, we instantiate the base policy with the conditional latent diffusion architecture used in CP-Composer~\citep{jiang2025zero}, enabling a controlled comparison with the strongest guidance-based baseline under the same generation interface. 
A VAE encoder $\mathcal{E}_{\phi}$ maps a peptide structure to residue-level latent variables $\mathbf{z}_{0}$, and a decoder $\mathcal{D}_{\psi}$ reconstructs full-atom structures from these latents. 
The forward diffusion process corrupts latents as
\begin{equation*}
q(\mathbf{z}_t \mid \mathbf{z}_0)
=
\mathcal{N}
\left(
\mathbf{z}_t;
\sqrt{\bar{\alpha}_t}\mathbf{z}_0,
(1-\bar{\alpha}_t)\mathbf{I}
\right),
\end{equation*}
and the reverse process is parameterized by 
$\boldsymbol{\epsilon}_{\theta}(\mathbf{z}_t,t,P,\mathcal{C})$.

At inference time, classifier-free guidance combines conditional and constraint-dropped predictions:
\begin{equation*}
\label{equ:cfg}
\tilde{\boldsymbol{\epsilon}}_\theta(\mathbf{z}_t,t,P,\mathcal{C})
=
(1+w)\boldsymbol{\epsilon}_\theta(\mathbf{z}_t,t,P,\mathcal{C})
-
w\boldsymbol{\epsilon}_\theta(\mathbf{z}_t,t,P,\emptyset),
\end{equation*}
where $w$ is the guidance strength and the constraint-dropped branch retains the target protein $P$. 
This conditional diffusion model defines the base policy. 
While CFG can steer samples at inference time, it does not directly change the learned generative distribution; GeoCycler instead aligns the policy parameters with macrocyclization rewards so that closure-consistent candidates become more likely under the generator. 
Additional details of the latent representation, constraint encoding, reward normalization, and replay-based fine-tuning are provided in Appendix~\ref{app:method_details}.

\subsection{Reward-Weighted Diffusion Alignment}

GeoCycler collects terminal samples from the current generator, decodes them into peptide candidates, scores them with topology-specific rewards, and uses the scores to reweight standard forward-diffusion denoising updates. 
This avoids differentiating through non-smooth constraint checks while keeping optimization close to the original diffusion objective.

\paragraph{Type-gated stair reward.}
Distance-based feedback is meaningful only when the required residue or linker types are present. 
We therefore define
\begin{equation*}
\label{eq:type_gated_reward}
R(\mathcal{M}\mid\mathcal{C})
=
\mathbb{I}[\mathcal{C}_{\mathrm{type}}(\mathcal{M})]\,
R_{\mathrm{dist}}(\mathcal{M}\mid\mathcal{C}_{\mathrm{dist}}),
\end{equation*}
where the type gate prevents distance rewards from being assigned to chemically incompatible anchors. 
The distance component is a bounded stair reward:
\begin{equation*}
\label{eq:stair_reward}
R_{\mathrm{dist}}
=
\mathbb{I}[\mathcal{C}_{\mathrm{dist}}(\mathcal{M})]
+
\kappa
\bigl(1-\mathbb{I}[\mathcal{C}_{\mathrm{dist}}(\mathcal{M})]\bigr)
\mathrm{shape}(\mathcal{M},\mathcal{C}_{\mathrm{dist}}),
\end{equation*}
where $\kappa\in(0,1)$ and $\mathrm{shape}(\cdot)\in[0,1]$ is a bounded function of topology-specific distance violations. 
Thus, exact proxy closure receives full reward, while near-feasible samples receive intermediate feedback.

\paragraph{Reward normalization.}
For a local population of $N_p$ samples, we compute rewards $R_i$, mean $\bar{R}$, and standard deviation $s$, and map each reward to a bounded weight:
\begin{equation*}
\label{eq:reward_weight}
r_i
=
\frac{1}{2}
+
\frac{1}{2}
\mathrm{clip}
\left(
\frac{R_i-\bar{R}}{s+\delta},
-1,
1
\right),
\quad
\bar{R}=\frac{1}{N_p}\sum_{j=1}^{N_p}R_j .
\end{equation*}
The clipping limits the effect of outliers while emphasizing samples that are more closure-consistent than the local population average.


\paragraph{Alignment objective.}
For each sampled terminal latent $\mathbf{z}_0\sim p_{\theta}(\cdot\mid P,\mathcal{C})$, we draw 
$t\sim\mathcal{U}[1,T]$ and $\boldsymbol{\epsilon}\sim\mathcal{N}(\mathbf{0},\mathbf{I})$, and construct
\begin{equation*}
\mathbf{z}_t
=
\sqrt{\bar{\alpha}_t}\mathbf{z}_0
+
\sqrt{1-\bar{\alpha}_t}\boldsymbol{\epsilon}.
\end{equation*}
The reward-weighted alignment loss is
\begin{equation}
\label{eq:align_loss}
\mathcal{L}_{\mathrm{align}}(\theta)
=
\mathbb{E}
\left[
r_i\,\lambda(t)
\left\|
\tilde{\boldsymbol{\epsilon}}_\theta(\mathbf{z}_t,t,P,\mathcal{C})
-
\boldsymbol{\epsilon}
\right\|_2^2
\right],
\end{equation}
where $\lambda(t)$ is the diffusion loss weight. 
We apply the objective to the same guidance-mixed predictor used at sampling time, aligning the deployed generator rather than training a separate scorer or relying on post hoc filtering. 

\subsection{Multi-Topology Training with Replay}

GeoCycler fine-tunes a single conditional generator across stapled, head-to-tail, disulfide, and bicyclic topologies. 
At each training iteration, we sample a cyclization strategy, construct its constraint specification $\mathcal{C}$, and update the shared denoising network using the corresponding topology-specific reward. 
This mixed-topology setup tests whether one reward-aligned generator can support multiple macrocyclization mechanisms rather than relying on separately tuned specialists.

To improve sample efficiency, we maintain a FIFO replay buffer
\begin{equation*}
\mathcal{D}=\{(P,\mathcal{C},\mathbf{z}_0,R)\}.
\end{equation*}
Mini-batches mix newly collected samples with replayed terminal latents. 
For each replayed $\mathbf{z}_0$, we resample $(t,\boldsymbol{\epsilon})$ on the fly and evaluate Eq.~\ref{eq:align_loss}, avoiding storage of full reverse trajectories. 
Replay increases the reuse of rare high-reward closure patterns and stabilizes reward-weighted fine-tuning.

\section{Experiments}
\label{sec:experiments}
\subsection{Experimental Setup}
\label{sec:experimental_setup}

\paragraph{Datasets.}
Following CP-Composer~\citep{jiang2025zero}, we use PepBench and ProtFrag for model development, and evaluate on the LNR benchmark~\citep{tsaban2022harnessing}. 
PepBench contains 4,157 training and 114 validation protein--peptide complexes, while ProtFrag provides 70,498 protein-fragment samples for structural pretraining~\citep{kong2024full}. 
The LNR test set contains 93 expert-curated protein--peptide complexes. 
Detailed filtering, splitting, and preprocessing procedures are provided in Appendix~\ref{app:datasets}.

\paragraph{Baselines.}
We compare GeoCycler with three representative generation baselines: PepGLAD~\citep{kong2024full}, an unconditional full-atom peptide diffusion model; w/EG~\citep{bao2022equivariant}, an energy-guided sampling baseline; and CP-Composer~\citep{jiang2025zero}, the strongest guidance-based cyclic peptide generator. 
For CP-Composer, we use the reported CFG setting $w=5$ in the main comparison and analyze guidance sensitivity in Section~\ref{sec:guidance_sensitivity}. 
An adapted DiffusionNFT-style positive/negative objective~\citep{zheng2025diffusionnft} is evaluated as a controlled alignment-objective ablation in Section~\ref{sec:mechanistic_ablation}.

\paragraph{Metrics.}
We report pass@5 success rate as the primary metric: a target is successful if at least one of five generated candidates satisfies the topology-specific geometric closure criterion. 
We also report pass@1 to measure single-sample yield. 
To monitor distributional drift after alignment, we report amino-acid KL divergence (AA-KL) and backbone-dihedral KL divergence (B-KL) relative to natural peptides. 
Topology-specific success criteria and training details are provided in Appendices~\ref{app:reward} and~\ref{app:training}.

\subsection{Main Results: Constraint-Conditioned 3D Generation}

Table~\ref{tab:main_results} compares GeoCycler with representative peptide generation and guidance-based baselines across four macrocyclization strategies. 
For GeoCycler, success rates are reported as the mean over three independent fine-tuning seeds, with standard deviations provided in Appendix~\ref{app:seed_robustness}. 
GeoCycler achieves the highest pass@5 success rate in all four settings.

\begin{table*}[!htp]
\centering
\caption{
\textbf{Target-conditioned cyclic peptide generation.}
Success rate is reported as pass@5. 
For GeoCycler, Succ. reports the mean over three independent fine-tuning seeds. 
AA-KL and B-KL are distributional fidelity diagnostics relative to natural peptides.
The highest success rate for each topology is shown in \textbf{bold}.
}
\label{tab:main_results}
\small
\setlength{\tabcolsep}{4.5pt}
\renewcommand{\arraystretch}{1.08}
\resizebox{\textwidth}{!}{
\begin{tabular}{l|ccc|ccc|ccc|ccc}
\toprule
\multirow{2}{*}{Method} 
& \multicolumn{3}{c|}{Stapled} 
& \multicolumn{3}{c|}{Head-to-tail}
& \multicolumn{3}{c|}{Disulfide}
& \multicolumn{3}{c}{Bicycle} \\
\cmidrule(lr){2-4}\cmidrule(lr){5-7}\cmidrule(lr){8-10}\cmidrule(lr){11-13}
& Succ. & AA-KL & B-KL 
& Succ. & AA-KL & B-KL 
& Succ. & AA-KL & B-KL 
& Succ. & AA-KL & B-KL \\
\midrule
PepGLAD 
& 22.80 & 0.1035 & 1.1401 
& 30.23 & 0.1052 & 1.1347
& 0.00 & 0.0808 & 1.1324
& 0.00 & 0.0838 & 1.1823 \\
w/EG 
& 25.41 & 0.0744 & 1.1821
& 61.63 & 0.0798 & 1.0891
& 0.00 & 0.0711 & 1.0891
& 0.00 & 0.0729 & 1.0968 \\
CP-Composer 
& 86.84 & 0.2827 & 1.0942
& 75.27 & 0.1206 & 1.0275
& 85.06 & 0.4855 & 0.9986
& 3.45 & 0.3303 & 1.1108 \\
\textbf{GeoCycler}
& \textbf{89.91} & 0.2852 & 1.0652
& \textbf{96.06} & 0.1261 & 1.0074
& \textbf{94.63} & 0.4209 & 1.0367
& \textbf{11.49} & 0.3520 & 1.1240 \\
\bottomrule
\end{tabular}
}
\end{table*}

The comparison highlights that distributional fidelity alone is not sufficient for macrocyclization. 
PepGLAD and w/EG obtain competitive or even lower KL values in several cases, but they fail to satisfy the harder closure constraints, especially for disulfide and bicyclic peptides. 
This suggests that cyclic peptide generation requires explicit alignment to topology-specific closure feasibility, rather than only preserving natural peptide statistics or applying generic energy guidance.
Compared with CP-Composer, GeoCycler delivers consistent gains across single-closure, side-chain, and multi-anchor cyclization settings. 
The 20.79-point improvement on head-to-tail cyclization shows that reward-weighted alignment effectively promotes terminal backbone closure. 
The gains on disulfide and bicyclic peptides further indicate that the type-gated reward helps coordinate residue prerequisites with closure geometry, enabling better satisfaction of coupled multi-edge constraints.

The fidelity diagnostics indicate that these gains are not driven by severe distributional drift. 
Across the four topologies, GeoCycler's AA-KL and B-KL remain in the same range as CP-Composer, with improved or comparable backbone-dihedral KL in stapled and head-to-tail settings. 
Overall, the results support the main premise of GeoCycler: reward-weighted training-time alignment increases the probability of closure-consistent samples while largely preserving the peptide distribution learned by the base generator.
For example, on head-to-tail cyclization, GeoCycler substantially improves success while maintaining similar AA-KL and slightly lower B-KL. 
We therefore interpret the main improvement as better allocation of probability mass toward closure-consistent samples under the same target-conditioned generation task.




\begin{wraptable}{l}{0.48\textwidth}
\vspace{-1.0em}
\centering
\caption{
\textbf{Single-shot success rate.}
Pass@1 measures the fraction of targets for which one sampled candidate satisfies the topology-specific geometric constraint.
}
\label{tab:pass1}
\small
\setlength{\tabcolsep}{5pt}
\renewcommand{\arraystretch}{1.05}
\begin{tabular}{lcc}
\toprule
Topology & CP-Composer & GeoCycler \\
\midrule
Stapled      & 40.79 & \textbf{42.11} \\
Head-to-tail & 38.71 & \textbf{46.24} \\
Disulfide    & 70.11 & \textbf{74.71} \\
Bicycle      & 0.00  & \textbf{3.45} \\
\bottomrule
\end{tabular}
\vspace{-1.0em}
\end{wraptable}

We also report pass@1 in Table~\ref{tab:pass1}. 
Unlike pass@5, which benefits from multiple attempts per target, pass@1 measures the single-trajectory yield of the generator. 
GeoCycler improves pass@1 across all topologies, including a gain from 38.71 to 46.24 on head-to-tail cyclization and from 0.00 to 3.45 on bicyclic generation. 
These results suggest that reward alignment increases the probability of generating closure-consistent candidates from a single sampled trajectory, rather than only improving success through repeated sampling.

\subsection{Guidance Sensitivity and Baseline Calibration}
\label{sec:guidance_sensitivity}

Classifier-free guidance provides a controlled probe of how a constraint-conditioned generator responds to inference-time steering. 
If macrocyclization feasibility were mainly recovered by post hoc guidance, increasing the guidance scale would be the dominant source of improvement. 
Instead, Table~\ref{tab:cfg_sensitivity} shows that both CP-Composer and GeoCycler achieve their best or tied-best performance at the calibrated setting $w=5$, while larger guidance does not consistently improve success.

Across matched guidance scales, GeoCycler outperforms CP-Composer in most settings, indicating an upward shift of the guidance-response profile. 
This suggests that type-gated reward alignment turns macrocyclization constraints into training-time alignment signals, rather than relying on CFG as the sole post hoc correction mechanism.

\begin{table*}[htp]
\centering
\caption{
\textbf{Guidance sensitivity of GeoCycler and CP-Composer.}
All values are pass@5 success rates. 
The reported setting is $w=5$, which is the best or tied-best scale for both methods across the evaluated topologies. 
The best value for each topology is shown in \textbf{bold}.
}
\label{tab:cfg_sensitivity}
\small
\setlength{\tabcolsep}{5.2pt}
\renewcommand{\arraystretch}{1.08}
\begin{tabular}{llcccc}
\toprule
Method & Topology & $w=0$ & $w=3$ & $w=5$ (reported) & $w=10$ \\
\midrule
\multirow{4}{*}{GeoCycler}
& Stapled      & 48.68 & 82.89 & \textbf{89.91} & 76.32 \\
& Head-to-tail & 46.24 & 83.87 & \textbf{96.06} & 86.02 \\
& Disulfide    & 10.75 & 74.71 & \textbf{94.63} & 86.21 \\
& Bicycle      & 3.45  & 6.90  & \textbf{11.49} & 3.45 \\
\midrule
\multirow{4}{*}{CP-Composer}
& Stapled      & 46.05 & 86.84 & 86.84 & 72.37 \\
& Head-to-tail & 30.11 & 72.04 & 75.27 & 75.27 \\
& Disulfide    & 9.95  & 73.56 & 85.06 & 85.06 \\
& Bicycle      & 0.00  & 3.45  & 3.45  & 3.45 \\
\bottomrule
\end{tabular}
\end{table*}

\subsection{Qualitative Structural Realization}

\begin{figure*}[t]
    \centering
    \includegraphics[width=0.98\linewidth]{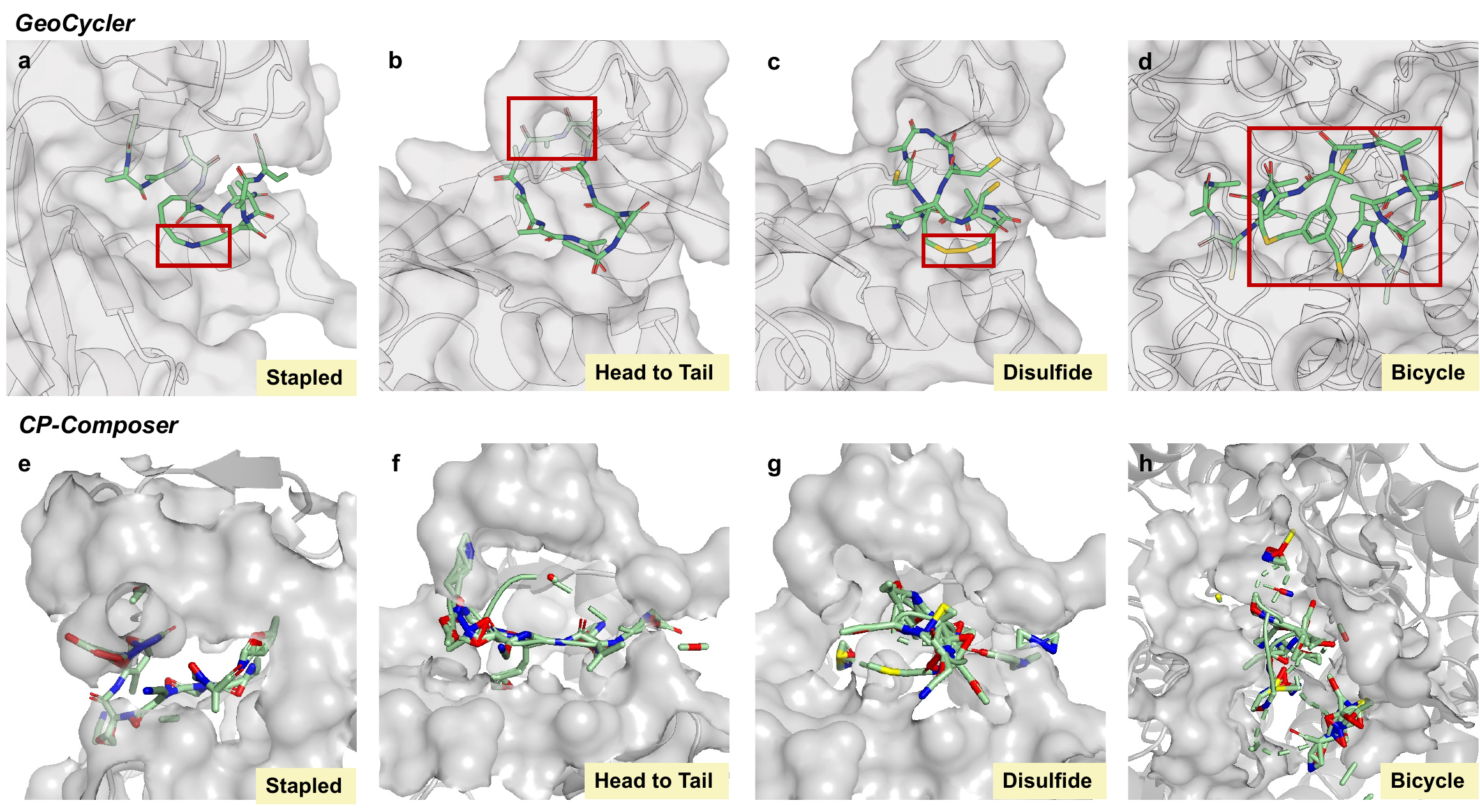}
    \caption{
    \textbf{Representative structural realizations across four macrocyclization topologies.}
    Panels \textbf{a--d} show GeoCycler samples, and panels \textbf{e--h} show CP-Composer samples. 
    Red boxes highlight the local regions associated with cyclization constraints.
    }
    \label{fig:case}
\end{figure*}

Figure~\ref{fig:case} provides representative structural realizations generated by GeoCycler and CP-Composer across stapled, head-to-tail, disulfide, and bicyclic settings. 
Beyond the binary closure metrics, these examples reveal how each model realizes the prescribed macrocyclization geometry in the local 3D environment. 
GeoCycler samples exhibit more coherent junction organization around the highlighted cyclization anchors, with smoother local backbone continuity, less visibly crowded anchor placement, and better accommodation within the protein pocket. 
In contrast, CP-Composer samples more frequently show distorted or crowded closure regions in the displayed cases, consistent with the difficulty of enforcing cyclic geometry through inference-time steering alone.

The bicyclic examples are particularly informative because this topology requires several closure constraints to be satisfied simultaneously. 
When one anchor is adjusted independently, the geometry around another closure edge can be perturbed, producing irregular multi-loop arrangements. 
GeoCycler's displayed sample forms a more organized multi-anchor geometry, suggesting that reward alignment encourages the generator to coordinate coupled closure constraints rather than correcting them only at the end of sampling. 
Together with the improvements in pass@5 and pass@1, these structural comparisons indicate that GeoCycler does not only increase threshold-based closure counts, but also tends to produce locally more plausible macrocyclization under the prescribed topology. 

As a secondary diagnostic, we further evaluate generated samples with an OpenMM-based post-hoc structural consistency screen, which checks whether samples retain topology-specific closure geometry after local relaxation and avoid severe steric clashes. The results are reported in Appendix~\ref{app:openmm_screen}.

\subsection{Mechanistic Ablations}
\label{sec:mechanistic_ablation}

We next isolate the contributions of GeoCycler's reward design, alignment objective, and training strategy. 
All ablations use the same base generator, data split, training budget, and pass@5 evaluation protocol. 

\paragraph{Reward design.}
GeoCycler's reward combines three ingredients: a prerequisite type gate, a topology-specific distance term, and bounded shaping for near-feasible samples. 
We compare the full reward with three variants. 
Only-Type keeps only the binary prerequisite type gate and removes geometric distance feedback. 
Only-Dist removes the type gate and applies distance feedback regardless of whether the required chemical anchors are present. 
No-Shaping keeps the type gate but removes the intermediate shaping term, leaving only binary closure success.
For H2T cyclization, the type gate is trivial in our formulation because no prerequisite residue pattern is required. 
We therefore set $G_{\mathrm{H2T}}=1$, and the H2T reward reduces to a shaped terminal-closure reward. 
Accordingly, the H2T entry of Only-Type is marked as not applicable, while the H2T Only-Dist variant is identical to the full shaped H2T reward and is not treated as an independent ablation.

\begin{wraptable}{l}{0.52\textwidth}
\vspace{-1.0em}
\centering
\caption{
\textbf{Ablation of reward design.}
}
\label{tab:ablation_reward}
\small
\setlength{\tabcolsep}{4.5pt}
\begin{tabular}{lccccc}
\toprule
Reward & Stapled & H2T & Disulfide & Bicycle \\
\midrule
Only-Type  & 86.84 & N/A & 91.95 & 10.34 \\
Only-Dist  & 85.53 & 96.06 & 93.10 & 6.70 \\
No-Shaping & 86.84 & 81.72 & 91.95 & 3.45 \\
GeoCycler  & \textbf{89.91} & \textbf{96.06} & \textbf{94.63} & \textbf{11.49} \\
\bottomrule
\end{tabular}
\end{wraptable}

Table~\ref{tab:ablation_reward} shows that neither the type gate nor the distance term alone is sufficient. 
Only-Type performs reasonably when residue prerequisites dominate, but lacks continuous geometric refinement, especially on head-to-tail cyclization. 
Only-Dist provides distance feedback without checking chemical compatibility, which is less effective for coupled bicyclic closure. 
No-Shaping further shows that binary closure rewards are too sparse, with clear drops on head-to-tail and bicycle generation. 
The full type-gated shaped reward achieves the best mean success rate, supporting the central design choice of using the type gate to define chemically meaningful anchors and bounded shaping to provide dense geometric credit assignment. 

\paragraph{Alignment objective.}
We next compare GeoCycler's positive-only reward-weighted alignment with positive/negative alternatives. 
DiffusionNFT-style Loss uses a combined positive/negative objective adapted to our cyclic peptide setting. 
Negative Loss keeps only the negative branch. 
Positive Loss is the GeoCycler objective, which reinforces higher-reward samples without explicitly fitting a heterogeneous negative set.

\begin{table}[htp]
\centering
\caption{
\textbf{Ablation of alignment objective.}
All variants use the same reward design, replay mechanism, mixed-topology setting, and training budget. 
}
\label{tab:ablation_loss}
\small
\setlength{\tabcolsep}{5pt}
\begin{tabular}{lccccc}
\toprule
Loss & Stapled & H2T & Disulfide & Bicycle & Mean \\
\midrule
DiffusionNFT-style Loss & 85.53 & 78.49 & 91.95 & 6.70  & 65.67 \\
Negative Loss           & 86.84 & 58.06 & 94.25 & 6.70  & 61.46 \\
Positive Loss           & \textbf{89.91} & \textbf{96.06} & \textbf{94.63} & \textbf{11.49} & \textbf{73.02} \\
\bottomrule
\end{tabular}
\end{table}

Table~\ref{tab:ablation_loss} shows that positive-only alignment achieves the best mean success rate. 
The gain is particularly large on head-to-tail cyclization, where negative-only alignment performs poorly, and on bicycle generation, where the feasible region is sparse. 
This supports our motivation for avoiding an explicit negative branch: low-reward cyclic peptide samples can fail for many unrelated reasons, including missing type prerequisites, incompatible anchors, poor closure distances, or conflicts among multiple closure edges. 
Such failures do not form a coherent repulsive direction. 
Positive-only alignment instead concentrates the update on feasible or near-feasible samples, which is better matched to sparse geometric constraint satisfaction.

\paragraph{Training strategy.}
Finally, we ablate two training-level choices. 
w/o Replay Buffer removes sample reuse while keeping the same reward and alignment objective. 
Strategy-specific trains separate models for each topology. 
Unified GeoCycler uses one mixed-topology model shared across all four cyclization strategies.

\begin{wraptable}{l}{0.65\textwidth}
\vspace{-1.0em}
\centering
\caption{
\textbf{Ablation of training strategy.}
}
\label{tab:ablation_training}
\small
\setlength{\tabcolsep}{4.5pt}
\begin{tabular}{lccccc}
\toprule
Variant & Stapled & H2T & Disulfide & Bicycle & Mean \\
\midrule
w/o Replay Buffer & 89.47 & 80.65 & 91.95 & 6.90  & 67.24 \\
Strategy-specific & 86.84 & 90.32 & 93.10 & 10.34 & 70.15 \\
Unified GeoCycler & \textbf{89.91} & \textbf{96.06} & \textbf{94.63} & \textbf{11.49} & \textbf{73.02} \\
\bottomrule
\end{tabular}
\end{wraptable}

Table~\ref{tab:ablation_training} shows that replay-based sample reuse improves mean success, with the largest gain on head-to-tail cyclization. 
This is consistent with the role of replay as a stabilizer for reward-weighted diffusion fine-tuning: rare high-reward closure patterns can be reused instead of being quickly lost in noisy on-policy sampling. 
The unified mixed-topology model also matches or outperforms strategy-specific fine-tuning across all topologies. 
This supports GeoCycler's framing as a single constraint-conditioned generator that can share geometric closure regularities across macrocyclization strategies, rather than a collection of separately tuned specialists.



\section{Conclusion, Scope, and Limitations}
\label{sec:conclusion}

We introduced \textbf{GeoCycler}, a reward-weighted diffusion alignment framework for constraint-conditioned cyclic peptide generation. 
By combining type-gated distance shaping, positive-only alignment, and replay-based stabilization, GeoCycler improves closure-consistent generation across stapled, head-to-tail, disulfide, and bicyclic settings on the LNR benchmark while preserving comparable amino-acid and backbone-dihedral statistics. 
Guidance-sensitivity analysis and ablations support that these gains arise from training-time alignment rather than inference-time steering alone.

GeoCycler uses lightweight topology-specific surrogate criteria to make sparse macrocyclization rewards tractable. 
Current evaluation focuses on prescribed geometric closure criteria, complemented by post-hoc relaxation-based diagnostics. 
Future work will extend the framework to richer chemical feasibility signals, noncanonical residues, realistic linker chemistries, and multi-objective design.

\bibliography{references}
\bibliographystyle{abbrvnat}
\appendix
\newpage
\section{Dataset and Benchmark Descriptions}
\label{app:datasets}
GeoCycler is trained and evaluated using supervised protein--peptide complexes and unsupervised protein fragments following prior peptide latent-diffusion protocols~\citep{kong2024full,jiang2025zero}.

\subsection{PepBench and LNR Benchmark}
The primary supervised dataset, \textbf{PepBench}, is constructed from all dimers extracted from the Protein Data Bank (PDB). We filter for complexes with receptors longer than 30 residues and ligands between 4 and 25 residues. To ensure non-redundancy, we remove duplicated complexes where both the receptor and peptide share over 90\% sequence identity, resulting in 6,105 non-redundant complexes.

For evaluation, we utilize the \textbf{Large Non-Redundant (LNR) dataset} curated by domain experts, which contains 93 protein-peptide complexes involving canonical amino acids \citep{tsaban2022harnessing}. To rigorously test cross-target generalization, we cluster the receptor sequences of the entire pool using a 40\% sequence identity threshold. Any clusters containing targets from the LNR test set are removed from the training pool. After removing LNR-overlapping clusters, the remaining pool is split into 4,157 training complexes and 114 validation complexes. The binding site for each complex is defined as receptor residues within 10~\text{\AA} of any peptide residue, measured by the distance between their $C_\beta$ coordinates.

\subsection{ProtFrag Unsupervised Dataset}
To facilitate the training of the variational autoencoder (VAE) and help the model internalize diverse structural motifs, we exploit \textbf{ProtFrag}, an unsupervised dataset comprising approximately 70,498 protein fragments derived from monomeric proteins in the PDB. Fragments are extracted based on stringent biophysical criteria to ensure stability and chemical balance:
\begin{itemize}[leftmargin=*,noitemsep]
    \item \textbf{Length and Composition}: Fragments must consist of 4 to 25 residues. No single amino acid can constitute more than 25\% of the fragment, and the composition of hydrophobic and charged residues is strictly balanced.
    \item \textbf{Biophysical Stability}: We filter for fragments with an instability index below 40. Additionally, fragments must exhibit a Buried Surface Area (BSA) above 400~\text{\AA}$^2$ and a relative BSA above 20\% when considering their surrounding residues in the parent protein, to retain compact and biophysically reasonable fragment-like structures.
\end{itemize}

\subsection{Data Summary}
The final composition of the datasets used for training and testing is summarized in Table~\ref{tab:dataset_stats}.

\begin{table}[h]
\centering
\caption{Statistics of the constructed datasets used for GeoCycler training and evaluation.}
\label{tab:dataset_stats}
\vskip 0.15in
\begin{small}
\begin{sc}
\begin{tabular}{lccc}
\toprule
Split & \#Entries & \#Clusters & Source \\
\midrule
PepBench (Training)   & 4,157  & 952 & PDB \\
PepBench (Validation) & 114    & 50  & PDB \\
PepBench (Test)       & 93     & 93  & LNR \\
ProtFrag (Unsupervised) & 70,498 & --  & PDB Monomers \\
\bottomrule
\end{tabular}
\end{sc}
\end{small}
\vskip -0.1in
\end{table}

\subsection{Evaluation Protocol on LNR}
\label{app:evaluation_protocol}

We use the same LNR source benchmark, but topology-specific eligibility filters lead to different numbers of valid evaluation targets. All methods are evaluated on the identical eligible target subset for each topology. 
For pass@5 evaluation, we generate five candidates per target and count the target as successful if at least one candidate satisfies the corresponding geometric closure criterion. 
For pass@1 evaluation, only the first sampled candidate is used. 
The same target set, sampling budget, and topology-specific success criteria are used for all compared methods.

\newpage

\section{Detailed Reward and Evaluation Criteria}
\label{app:reward}

\subsection{Surrogate Closure Criteria}
\label{app:topology_definitions}

GeoCycler uses topology-specific surrogate closure criteria to provide lightweight and stable reward signals for constraint-conditioned cyclic peptide generation. 
Each macrocyclization topology is represented by two components: a discrete type gate, which checks whether the required residue or linker pattern is present, and a continuous geometric criterion, which evaluates whether the corresponding anchor-proxy geometry satisfies the prescribed closure condition.


For a generated peptide $M=(X,S)$ and a topology $\tau$, we define a topology-specific anchor-proxy selector
\begin{equation*}
    A_{\tau}(M)=\{a_1^\tau,\ldots,a_{m_\tau}^\tau\},
\end{equation*}
where $A_{\tau}(M)$ denotes the set of anchor-proxy coordinates used to evaluate the corresponding closure geometry. 
The selector follows the same constraint interface for all compared methods, ensuring that the reward computation and evaluation protocol are applied consistently across baselines and GeoCycler.

We consider four representative macrocyclization topologies:

\begin{itemize}[leftmargin=*,noitemsep]
    \item \textbf{Stapled peptide.}
    A side-chain stapling condition is represented by a Lysine--acidic-residue motif, where a Lysine ($K$) at position $i$ is paired with either Aspartic Acid ($D$) or Glutamic Acid ($E$) at position $j\in\{i+3,i+4\}$. 
    The geometric criterion evaluates whether the corresponding anchor-proxy distance lies within an admissible closure window.

    \item \textbf{Head-to-tail cyclization.}
    Head-to-tail cyclization imposes a terminal closure condition between the two ends of the peptide chain. 
    This topology has no nontrivial residue-type prerequisite in our formulation, so the type gate is always satisfied and the reward is determined by the terminal anchor-proxy distance.

    \item \textbf{Disulfide cyclization.}
    Disulfide-style cyclization requires at least two Cysteine ($C$) residues. 
    Since the anchor positions are not fixed in advance, the geometric criterion searches over candidate cysteine pairs and evaluates the closest feasible anchor-proxy pair.

    \item \textbf{Bicyclic peptide.}
    Bicyclic peptides require coupled multi-point closure. 
    In our setting, the type gate requires a peptide of length at least $13$ with exactly three or four Cysteine ($C$) residues. 
    The geometric criterion searches over candidate cysteine triplets and evaluates whether their anchor-proxy distances form a feasible triangle window.
\end{itemize}

These topology definitions are used consistently for reward computation during alignment and for the topology-specific geometric success criteria in the main evaluation.

\subsection{General Type-Gated Stair Reward}
\label{app:general_reward}

For each topology $\tau$, GeoCycler defines a binary type gate $G_{\tau}(M)\in\{0,1\}$ and a nonnegative geometric violation $e_{\tau}(M,C)\geq 0$. 
The condition $e_{\tau}(M,C)=0$ indicates that the generated peptide satisfies the topology-specific surrogate closure criterion. 
The reward is factorized into the type gate and a bounded distance-shaped stair reward:
\begin{equation*}
\label{eq:app_reward_general}
\hat R_{\tau}(M|C)
=
G_{\tau}(M)
\left[
\mathbb{I}\!\left[e_{\tau}(M,C)=0\right]
+
\kappa
\left(
1-\mathbb{I}\!\left[e_{\tau}(M,C)=0\right]
\right)
\operatorname{shape}_{\tau}\!\left(e_{\tau}(M,C)\right)
\right],
\end{equation*}
where $\kappa\in(0,1)$ controls the magnitude of intermediate shaping feedback. 
In our experiments, we set $\kappa=0.1$.

The shaping function is defined as a bounded squared-linear decay:
\begin{equation}
\label{eq:app_shape}
\operatorname{shape}_{\tau}(e)
=
\left(
1-\frac{\min(e,e_{\max}^{\tau})}{e_{\max}^{\tau}}
\right)^2,
\end{equation}
where $e_{\max}^{\tau}$ is a topology-specific shaping window. 
This formulation assigns full reward to samples that satisfy the surrogate closure criterion and provides bounded intermediate feedback to near-feasible samples. 
The bounded form prevents outlier geometries from dominating reward-weighted updates while preserving a useful learning signal under sparse closure constraints.

\subsection{Topology-Specific Geometric Violations}
\label{app:topology_rewards}

We next define the type gate $G_{\tau}$ and geometric violation $e_{\tau}$ for each topology. 
For compactness, all distance calculations are expressed in terms of the topology-specific anchor-proxy coordinates selected by $A_{\tau}(M)$.

\paragraph{Stapled peptide.}
The type gate is satisfied when the generated sequence contains a $K$--$(D/E)$ motif at an allowed offset:
\begin{equation*}
G_{\mathrm{staple}}(M)
=
\mathbb{I}
\left[
\exists i,\ j\in\{i+3,i+4\}
\text{ such that }
S_i=K,\ S_j\in\{D,E\}
\right].
\end{equation*}
For each valid motif pair, let $d_{ij}^{\mathrm{staple}}$ denote the corresponding anchor-proxy distance. 
The violation for a candidate pair is
\begin{equation*}
e_{ij}^{\mathrm{staple}}
=
\max\left(0,\ r_{\min}^{\mathrm{staple}}-d_{ij}^{\mathrm{staple}},\ 
d_{ij}^{\mathrm{staple}}-r_{\max}^{\mathrm{staple}}\right),
\end{equation*}
and the final violation is the smallest violation over all valid motif pairs:
\begin{equation*}
e_{\mathrm{staple}}(M,C)
=
\min_{(i,j)\in\mathcal{P}_{\mathrm{staple}}}
e_{ij}^{\mathrm{staple}}.
\end{equation*}
We use $e_{\max}^{\mathrm{staple}}=15.0$~\AA{} in Eq.~\ref{eq:app_shape}.

\paragraph{Head-to-tail cyclization.}
Head-to-tail cyclization has no nontrivial residue-type prerequisite in our formulation:
\begin{equation*}
G_{\mathrm{H2T}}(M)=1.
\end{equation*}
Let $d^{\mathrm{H2T}}$ denote the terminal anchor-proxy distance. 
The violation is defined relative to the prescribed terminal closure threshold:
\begin{equation*}
e_{\mathrm{H2T}}(M,C)
=
\max\left(0,\ d^{\mathrm{H2T}}-r_{\mathrm{H2T}}\right).
\end{equation*}
We use $e_{\max}^{\mathrm{H2T}}=10.0$~\AA{} in Eq.~\ref{eq:app_shape}.

\paragraph{Disulfide cyclization.}
The type gate is satisfied when the generated sequence contains at least two Cysteine residues:
\begin{equation*}
G_{\mathrm{SS}}(M)
=
\mathbb{I}\left[|\mathcal{I}_{\mathrm{Cys}}|\geq 2\right],
\end{equation*}
where $\mathcal{I}_{\mathrm{Cys}}=\{i:S_i=C\}$.
For each candidate cysteine pair $(i,j)$, let $d_{ij}^{\mathrm{SS}}$ denote the corresponding anchor-proxy distance. 
The pairwise violation is
\begin{equation*}
e_{ij}^{\mathrm{SS}}
=
\max\left(0,\ d_{ij}^{\mathrm{SS}}-r_{\mathrm{SS}}\right),
\end{equation*}
and the final violation is the smallest violation over all cysteine pairs:
\begin{equation*}
e_{\mathrm{SS}}(M,C)
=
\min_{i,j\in\mathcal{I}_{\mathrm{Cys}},\,i<j}
e_{ij}^{\mathrm{SS}}.
\end{equation*}
We use $e_{\max}^{\mathrm{SS}}=10.0$~\AA{} in Eq.~\ref{eq:app_shape}.

\paragraph{Bicyclic peptide.}
The type gate is satisfied when the sequence length and cysteine count match the prescribed bicyclic setting:
\begin{equation*}
G_{\mathrm{bicycle}}(M)
=
\mathbb{I}
\left[
L\geq 13
\ \land\
|\mathcal{I}_{\mathrm{Cys}}|\in\{3,4\}
\right].
\end{equation*}
For each candidate cysteine triplet $(i,j,k)$, we compute the three anchor-proxy distances
$d_{ij}^{\mathrm{bi}}$, $d_{jk}^{\mathrm{bi}}$, and $d_{ki}^{\mathrm{bi}}$. 
A single edge is considered feasible if its distance lies in the prescribed triangle window $(r_{\min}^{\mathrm{bi}}, r_{\max}^{\mathrm{bi}})$. 
The edge violation is
\begin{equation*}
e_{\mathrm{edge}}(d)
=
\begin{cases}
0, & r_{\min}^{\mathrm{bi}} < d < r_{\max}^{\mathrm{bi}},\\
\min\left(|d-r_{\min}^{\mathrm{bi}}|,\ |d-r_{\max}^{\mathrm{bi}}|\right), & \text{otherwise}.
\end{cases}
\end{equation*}
The triplet-level violation is the maximum violation among the three edges:
\begin{equation*}
e_{ijk}^{\mathrm{bi}}
=
\max\left(
e_{\mathrm{edge}}(d_{ij}^{\mathrm{bi}}),
e_{\mathrm{edge}}(d_{jk}^{\mathrm{bi}}),
e_{\mathrm{edge}}(d_{ki}^{\mathrm{bi}})
\right).
\end{equation*}
The final bicyclic violation is the best triplet violation:
\begin{equation*}
e_{\mathrm{bicycle}}(M,C)
=
\min_{i,j,k\in\mathcal{I}_{\mathrm{Cys}},\,i<j<k}
e_{ijk}^{\mathrm{bi}}.
\end{equation*}
Following the evaluation setting, we use $(r_{\min}^{\mathrm{bi}},r_{\max}^{\mathrm{bi}})=(6.0,10.0)$~\AA{} and $e_{\max}^{\mathrm{bicycle}}=8.0$~\AA{}.

\subsection{Summary of Reward Hyperparameters}
\label{app:reward_hyperparameters}

Table~\ref{tab:reward_summary} summarizes the type gates, surrogate geometric criteria, and shaping windows used in GeoCycler. 
The topology-specific thresholds define the lightweight surrogate closure criteria used for reward computation and pass@5 evaluation, while the shaping windows $e_{\max}^{\tau}$ determine how near-feasible violations are mapped to bounded intermediate rewards.

\begin{table}[t]
\centering
\caption{
Summary of topology-specific surrogate reward criteria.
The anchor-proxy selector is applied consistently across GeoCycler and all compared methods under the same evaluation protocol.
}
\label{tab:reward_summary}
\resizebox{\linewidth}{!}{
\begin{tabular}{llll}
\toprule
Topology & Type gate & Surrogate geometric criterion & Shaping window \\
\midrule
Stapled 
& $K$--$(D/E)$ motif at allowed offset 
& Pairwise anchor-proxy distance within $[r_{\min}^{\mathrm{staple}},r_{\max}^{\mathrm{staple}}]$ 
& $e_{\max}=15.0$~\AA{} \\
Head-to-tail 
& None, $G_{\mathrm{H2T}}=1$ 
& Terminal anchor-proxy distance below $r_{\mathrm{H2T}}$ 
& $e_{\max}=10.0$~\AA{} \\
Disulfide 
& At least two Cysteine residues 
& Closest cysteine-pair anchor-proxy distance below $r_{\mathrm{SS}}$ 
& $e_{\max}=10.0$~\AA{} \\
Bicycle 
& $L\geq 13$ and $|\mathcal{I}_{\mathrm{Cys}}|\in\{3,4\}$ 
& Best cysteine-triplet anchor-proxy triangle in $(6.0,10.0)$~\AA{} 
& $e_{\max}=8.0$~\AA{} \\
\bottomrule
\end{tabular}
}
\end{table}

\newpage

\section{Additional Method Details}
\label{app:method_details}

\subsection{Latent Representation and Base Diffusion Policy}
\label{app:latent_policy}

GeoCycler uses CP-Composer as the base conditional latent diffusion policy. 
Given a peptide $\mathcal{M}=(\mathbf{X},\mathbf{S})$, the VAE encoder $\mathcal{E}_{\phi}$ maps the peptide into residue-level latent variables
\begin{equation*}
\mathbf{z}
=
\{(\mathbf{z}^{\mathrm{inv}}_i,\mathbf{z}^{\mathrm{eq}}_i)\}_{i=1}^{L},
\end{equation*}
where $\mathbf{z}^{\mathrm{inv}}_i\in\mathbb{R}^{d}$ captures $E(3)$-invariant residue features and 
$\mathbf{z}^{\mathrm{eq}}_i\in\mathbb{R}^{3}$ preserves equivariant geometric information. 
The decoder $\mathcal{D}_{\psi}$ reconstructs full-atom peptide structures from these latents.

The forward diffusion process corrupts clean latents $\mathbf{z}_0$ as
\begin{equation*}
q(\mathbf{z}_t\mid\mathbf{z}_0)
=
\mathcal{N}
\left(
\mathbf{z}_t;
\sqrt{\bar{\alpha}_t}\mathbf{z}_0,
(1-\bar{\alpha}_t)\mathbf{I}
\right),
\end{equation*}
where $\bar{\alpha}_t$ follows a fixed noise schedule. 
The reverse process is parameterized by a noise-prediction network
\begin{equation*}
\boldsymbol{\epsilon}_{\theta}(\mathbf{z}_t,t,P,\mathcal{C}).
\end{equation*}
During classifier-free guidance, conditional and constraint-dropped predictions are combined as
\begin{equation*}
\tilde{\boldsymbol{\epsilon}}_\theta(\mathbf{z}_t,t,P,\mathcal{C})
=
(1+w)\boldsymbol{\epsilon}_\theta(\mathbf{z}_t,t,P,\mathcal{C})
-
w\boldsymbol{\epsilon}_\theta(\mathbf{z}_t,t,P,\emptyset),
\end{equation*}
where $w$ is the guidance scale. 
The constraint-dropped branch retains the target protein $P$ while removing the cyclization condition.

\subsection{Constraint Encoding}
\label{app:constraint_encoding}

The macrocyclization condition is represented as 
$\mathcal{C}=(\mathcal{C}_{\mathrm{type}},\mathcal{C}_{\mathrm{dist}})$. 
The type component $\mathcal{C}_{\mathrm{type}}$ is encoded as residue-level features indicating prerequisite residue or linker requirements. 
The distance component $\mathcal{C}_{\mathrm{dist}}$ is encoded as edge-level geometric features, using radial basis encodings of the prescribed closure distances. 
This follows the constraint-conditioned design protocol of CP-Composer, while GeoCycler modifies the training objective through reward-weighted alignment rather than changing the base conditioning interface.

\subsection{Reward Normalization}
\label{app:reward_normalization}

For a local population $\mathcal{P}$ of size $N_p$, each generated sample receives a scalar topology-specific reward $R_i$. 
We normalize rewards within the local population to reduce scale sensitivity:
\begin{equation*}
a_i = R_i-\bar{R},
\qquad
\bar{R}=\frac{1}{N_p}\sum_{j=1}^{N_p}R_j,
\end{equation*}
\begin{equation*}
s=
\sqrt{
\frac{1}{N_p}
\sum_{j=1}^{N_p}
(R_j-\bar{R})^2
}.
\end{equation*}
The normalized advantage is mapped to a bounded optimization weight:
\begin{equation*}
r_i
=
\frac{1}{2}
+
\frac{1}{2}
\mathrm{clip}
\left(
\frac{a_i}{s+\delta},
-1,
1
\right),
\end{equation*}
where $\delta$ is a small numerical constant. 
The bounded mapping limits the influence of outliers and stabilizes fine-tuning when rewards are sparse.

\subsection{Reward-Weighted Forward-Diffusion Objective}
\label{app:forward_diffusion_alignment}

For each condition $(P,\mathcal{C})$, GeoCycler first samples a terminal latent 
$\mathbf{z}_0\sim p_{\theta}(\cdot\mid P,\mathcal{C})$ through the full denoising process and decodes it into a peptide candidate. 
The decoded candidate is scored using the topology-specific reward. 
We then sample $t\sim\mathcal{U}[1,T]$ and $\boldsymbol{\epsilon}\sim\mathcal{N}(\mathbf{0},\mathbf{I})$, and construct
\begin{equation*}
\mathbf{z}_t
=
\sqrt{\bar{\alpha}_t}\mathbf{z}_0
+
\sqrt{1-\bar{\alpha}_t}\boldsymbol{\epsilon}.
\end{equation*}
The alignment objective is
\begin{equation*}
\mathcal{L}_{\mathrm{align}}(\theta)
=
\mathbb{E}
\left[
r_i\,\lambda(t)
\left\|
\tilde{\boldsymbol{\epsilon}}_\theta(\mathbf{z}_t,t,P,\mathcal{C})
-
\boldsymbol{\epsilon}
\right\|_2^2
\right],
\end{equation*}
where $\lambda(t)$ is the diffusion loss weight. 
The reward weight $r_i$ is treated as a sample weight. 
Thus, the model reinforces generated latents that decode into more closure-consistent candidates while keeping the update close to the original denoising objective.

\subsection{Mixed-Topology Fine-Tuning and Replay}
\label{app:mixed_topology_replay}

GeoCycler fine-tunes a single conditional generator across stapled, head-to-tail, disulfide, and bicyclic topologies. 
At each training iteration, one topology is sampled, its constraint specification $\mathcal{C}$ is constructed, and the shared denoising network is updated using the corresponding topology-specific reward.

To improve sample efficiency, we use a first-in-first-out replay buffer:
\begin{equation*}
\mathcal{D}
=
\{(P,\mathcal{C},\mathbf{z}_0,R)\}.
\end{equation*}
Mini-batches are formed by mixing newly collected samples with replayed terminal latents. 
For each replayed $\mathbf{z}_0$, we resample $(t,\boldsymbol{\epsilon})$ on the fly and evaluate the same reward-weighted forward-diffusion objective. 
This avoids storing full reverse denoising trajectories while increasing reuse of rare high-reward closure patterns.

\newpage
\section{Training Implementation Details}
\label{app:training}
All policy-alignment experiments were conducted on NVIDIA H800 GPUs to ensure high-throughput structural optimization. Regarding the optimization strategy, the framework utilizes the AdamW optimizer with a learning rate of $5.0 \times 10^{-6}$ and a gradient clipping threshold of $6.0$, facilitating stable convergence during the initial stages of reward-driven manifold exploration. To mitigate the high variance of reinforcement learning in sparse geometric manifolds and enhance sample efficiency, we implement an off-policy stabilization mechanism with a replay sample ratio of $5.0$; this ensures that each newly collected trajectory is augmented with five historical samples retrieved from the replay buffer, providing a more robust and diverse optimization signal during the weight update. For sampling and collection, we use 100-step solver for both training rollouts and evaluation; specifically, in training stage, $K=10$ trajectories are sampled per condition to estimate a group-centered reward baseline, which serves to reduce the variance of the policy gradient and accelerate the internalization of cyclization geometry. To accommodate the variable sequence lengths inherent in de novo peptide design, we implement a dynamic batching strategy with a quadratic complexity bound of $6000$ (calculated as $n^2$ where $n$ is the number of residues), ensuring robust memory management and consistent throughput throughout the mixed-topology fine-tuning process.

\newpage
\section{Seed Robustness of GeoCycler}
\label{app:seed_robustness}

To assess the robustness of GeoCycler, we repeat the full reward-alignment procedure with three independent fine-tuning seeds under the same data split, training protocol, sampling budget, and pass@5 evaluation setting. 
Table~\ref{tab:seed_robustness} reports the mean and standard deviation of GeoCycler across the three runs. 
For reference, we also include the CP-Composer result used in the main comparison.

\begin{table}[h]
\centering
\caption{
\textbf{Seed robustness of GeoCycler.}
GeoCycler is evaluated over three independent fine-tuning seeds. 
CP-Composer is shown as the reference guidance-based baseline used in the main comparison. 
All values are pass@5 success rates.
}
\label{tab:seed_robustness}
\vskip 0.1in
\begin{small}
\begin{tabular}{lcc}
\toprule
Topology & CP-Composer & GeoCycler, mean $\pm$ std \\
\midrule
Stapled      & 86.84 & $\mathbf{89.91 \pm 0.76}$ \\
Head-to-tail & 75.27 & $\mathbf{96.06 \pm 1.64}$ \\
Disulfide    & 85.06 & $\mathbf{94.63 \pm 0.66}$ \\
Bicycle      & 3.45  & $\mathbf{11.49 \pm 1.99}$ \\
\bottomrule
\end{tabular}
\end{small}
\vskip -0.1in
\end{table}

GeoCycler remains above the CP-Composer reference baseline across all four cyclization topologies. 
The variation across seeds is small for stapled, head-to-tail, and disulfide cyclization, and remains moderate for the more challenging bicyclic setting. 
These results indicate that the main improvements are not attributable to a single favorable fine-tuning run.

\newpage
\section{OpenMM-based Post-hoc Structural Consistency Screen}
\label{app:openmm_screen}

\paragraph{Motivation.}
The main evaluation in this work follows topology-specific geometric closure criteria, which are lightweight enough to support reward-based diffusion alignment. These criteria are useful for measuring whether generated candidates satisfy the prescribed macrocyclization geometry, but they do not fully assess whether the structures remain locally consistent after molecular-mechanics relaxation or whether they contain severe steric artifacts. To provide an additional diagnostic beyond the training-time proxy reward, we conduct an OpenMM-based post-hoc structural consistency screen.

The goal of this analysis is: to test whether a generated candidate can retain the intended topology-specific closure geometry after local OpenMM relaxation while avoiding severe steric clashes.

\paragraph{Protocol.}
For each generated peptide--protein complex, we apply the same OpenMM-based local relaxation procedure to samples from CP-Composer and GeoCycler. The relaxation procedure is used only for post-hoc evaluation and is not used during GeoCycler training, reward computation, or sample selection.

After relaxation, each generated candidate is evaluated by two filters:
\begin{enumerate}
    \item \textbf{Retained closure geometry.} The candidate is checked for whether it still satisfies the topology-specific reactive-center or closure-geometry criterion after relaxation.
    \item \textbf{Severe-clash filtering.} The candidate is checked for whether it avoids severe steric overlaps according to the predefined clash criterion.
\end{enumerate}

For each topology and method, we report three quantities:
\begin{itemize}
    \item \textbf{Retained}: the fraction of evaluated targets for which the generated samples retain the topology-specific closure geometry after relaxation;
    \item \textbf{Clash-free}: the fraction of evaluated targets for which the generated samples pass the severe-clash filter;
    \item \textbf{Success}: the pass@5 success rate under the combined criterion, where a target is counted as successful only if at least one of five generated candidates both retains the closure geometry and passes the severe-clash filter.
\end{itemize}

\paragraph{Results.}
Table~\ref{tab:openmm_screen} reports the OpenMM-screened post-hoc structural consistency results across four macrocyclization topologies. Compared with CP-Composer, GeoCycler improves the final retained-and-clash-free success rate on stapled, head-to-tail, and disulfide cyclization. 

\begin{table}[ht]
\centering
\caption{
\textbf{OpenMM-based post-hoc structural consistency screen.}
After OpenMM-based local relaxation, a target is counted as successful only if at least one of five generated candidates both retains the topology-specific reactive-center or closure geometry and passes the severe-clash filter. This screen is used as a secondary diagnostic.
}
\label{tab:openmm_screen}
\begin{tabular}{lcccccc}
\toprule
\multirow{2}{*}{Topology}
& \multicolumn{3}{c}{CP-Composer}
& \multicolumn{3}{c}{GeoCycler} \\
\cmidrule(lr){2-4}
\cmidrule(lr){5-7}
& Retained & Clash-free & Success
& Retained & Clash-free & Success \\
\midrule
Stapled      & 69.89\% & 38.71\% & 10.75\% & \textbf{73.12\%} & \textbf{39.78\%} & \textbf{12.90\%} \\
Head-to-tail & 68.82\% & 41.94\% & 15.05\% & \textbf{92.47\%} & \textbf{43.01\%} & \textbf{20.43\%} \\
Disulfide    & \textbf{83.87\%} &  5.38\% &  3.23\% & 82.80\% & \textbf{18.30\%} &  \textbf{9.68\%} \\
Bicycle      &  8.60\% & 50.54\% &  0.00\% &  \textbf{9.68\% }& \textbf{53.76\%} &  0.00\% \\
\bottomrule
\end{tabular}
\end{table}

\paragraph{Interpretation.}
The OpenMM-screened results are consistent with the main conclusion that training-time reward alignment improves the probability of generating closure-consistent candidates that remain more robust under post-hoc structural filtering. The improvement is most evident for head-to-tail cyclization, where GeoCycler substantially increases retained closure after relaxation and improves the combined success rate from 15.05\% to 20.43\%. For disulfide cyclization, GeoCycler achieves a much higher clash-free rate and improves the combined success rate from 3.23\% to 9.68\%, although the retained-closure rate is slightly lower than CP-Composer. For stapled peptides, GeoCycler provides a modest improvement in both retained closure and final success.

The bicyclic setting remains difficult under this stricter screen. Although GeoCycler slightly improves retained geometry and clash-free rates, neither method achieves a retained-and-clash-free pass@5 success under the combined criterion. This result suggests that coupled multi-anchor macrocyclization remains a challenging case.



\end{document}